\documentclass[]{interact}
\usepackage{epstopdf}
\usepackage[caption=false]{subfig}
\usepackage{amssymb}
\usepackage{bbold}
\usepackage{mathtools}
\usepackage{bm}
\usepackage{graphicx}
\usepackage{color}

\begin{document}
\title{Cyclone-anticyclone asymmetry in rotating thin fluid layers}

\author{
  \name{G. Boffetta\textsuperscript{a}
    \thanks{CONTACT G. Boffetta Email: guido.boffetta@unito.it},
    F. Toselli\textsuperscript{a},
    M. Manfrin\textsuperscript{a} and
    S. Musacchio\textsuperscript{a}}
  \affil{\textsuperscript{a}
    Dipartimento di Fisica and INFN, Universit\`a di Torino,
    via P. Giuria 1, 10125 Torino, Italy}}

\maketitle

\begin{abstract}
We report of a series of laboratory experiments and numerical simulations 
of freely-decaying rotating turbulent flows confined in domains with variable 
height. We show that the vertical confinement has important effects on the
formation of large-scale columnar vortices, the hallmark of rotating 
turbulence, and in particular delays the development of the 
cyclone-anticyclone asymmetry.
We compare the experimental and numerical results face-to-face, showing
the robustness of the results.
\end{abstract}


%%%%%%%%%%%%%%%%%%%%%%%%%%%%%%%%%%%%%%%%%%%%%%%%%%%%%%%%%%%%%%%%%
\section{Introduction}
\label{sec1}

A distinctive feature of turbulent rotating flows is the spontaneous
formation of coherent columnar vortices aligned in the direction of the rotation axis.
The presence of these long-living, quasi-two-dimensional structures
have been observed both in experiments
\cite{Hopfinger1982,Longhetto2002,Staplehurst2008,Moisy2011}
and in numerical simulations
\cite{Bartello1994,Yeung1998,Smith1999,Yoshimatsu2011,Biferale2016}.
The mechanisms which cause their formation,
in particular concerning the interplay between inertial waves
and nonlinear triadic interactions, 
have been subject of intense studies
(for a recent review see, e.g., Ref.~\cite{Godeferd2015}).

Remarkably, most of these vortices are always co-rotating with the flow, 
i.e., they are cyclones.
The predominance of cyclones over anticyclones have been reported 
and investigated in a large number of numerical and experimental studies, 
both in freely decaying turbulence
~\cite{Bartello1994,Bourouiba2007,VanBokhoven2008,Morize2005,Morize2006,Praud2006,Moisy2011}
and forced turbulence
~\cite{Godeferd1999,Smith2005,Gallet2014,Biferale2016}. 
It has been observed also in atmospheric measurements \cite{Cho2001,Hakim2005}
and in rotating thermal convection \cite{cheng2015laboratory,guervilly2014large,
vorobieff1998vortex}.
The symmetry-breaking is typically quantified in terms of the
skewness $S_\omega = \langle \omega_z^3 \rangle/\langle \omega_z^2 \rangle^{3/2}$
of the vorticity $\omega_z$ in the direction of the rotation
vector ${\bm \Omega} = \Omega {\bm e}_z$.  
Other indicators have been recently introduced, including 
third-order two-point velocity correlation functions~\cite{Gallet2014},
the skewness of the azimuthal velocity increments~\cite{Deusebio2014}
and the alignment statistics between vorticity and the rotation vector~\cite{Naso2015}.

Two types of arguments have been proposed to explain this phenomenon.
First, cyclones have a larger vortex stretching $(2\Omega + \omega_z) \partial u_z /\partial z$
in a rotating flow with given vertical strain $\partial u_z / \partial z$.
As a consequence, an isotropic turbulent flow suddenly put into rotation
develops a positive skewness $S_\omega$~\cite{Gence2001}.
The second type of explanations is based on the Rayleigh stability criterion,
which shows that anticyclonic vortices
are more subject to centrifugal instabilities~\cite{Bartello1994,Sreenivasan2008}.

Previous studies have shown that the asymmetry is strongly dependent on the Rossby number $Ro$. 
In particular, it is maximum for $Ro$ of order unity~\cite{Bourouiba2007}.
In decaying rotating flows the skewness $S_\omega$ grows in time
as $Ro$ decreases from an initial large value~\cite{Morize2005,Moisy2011,Naso2015}.
A recovery of the symmetry has been observed in the late stage of
the decay, when $Ro \ll 1$~\cite{Morize2005,Moisy2011}.
Much less is known about the dependence of the asymmetry on the the height $H$
of the fluid in the direction of the rotation axis,
because this phenomenon is typically studied in domains with aspect ratio of order unity.
Recently, it has been shown that the confinement of the flow in a thin layer
causes a reduction of the asymmetry in forced rotating turbulence~\cite{Deusebio2014}.

The aim of this paper is to investigate, by means of experiments and numerical
simulations of freely decaying rotating turbulence how the cyclone-anticyclone
asymmetry is affected by the thickness of the flow.
This issue is closely connected to the puzzling relation
between rotation and two-dimensionalization in turbulence. 
On the one hand, it is well known that rotation induces a 
two-dimensionalization of turbulent flow,
which becomes almost invariant along the rotation vector ${\bm \Omega}$. 
On the other hand, the Coriolis force affects the dynamics of the velocity field only
if the latter has non-vanishing gradients in the direction of ${\bm \Omega}$.
In particular, in a perfectly two-dimensional (2D) flow the effects of rotation disappear
because the Coriolis force is canceled by pressure gradients.
Considering that the reduction of the thickness $H$ of the layer
enhances the two-dimensionalization of the flow~\cite{Deusebio2014}, 
we expect that also the cyclone-anticyclone asymmetry should be suppressed by the confinement.

In our study, the comparison of experiments and numerical simulations
is not intended to reproduce exactly the same physical setup.
Our aim is to compare two systems with structural differences related to their boundary conditions. 
In the experiment, the turbulent flow is subject to friction with
the bottom wall of the tank, which causes the development of an Ekman layer. 
In the numerics the boundary conditions are periodic in all directions and
the bottom friction is absent.
In the numerical simulations the large-scale energy transfer induced
by rotation and vertical confinement~\cite{Deusebio2014} 
eventally leads to the phenomenon of spectral condensation at the horizontal scale of box. 
In the experiments this phenomenon does not occur because
the turbulent flow is surrounded by still fluid and the diameter of the tank
is much larger than the typical size of the vortices generated by the comb.
Despite these differences, we show that the effects of the vertical confinement on the cyclone-anticyclone
asymmetry is similar: it causes a retardation of the growth of $S_\omega$.

%%%%%%%%%%%%%%%%%%%%%%%%%%%%%%%%%%%%%%%%%%%%%%%%%%%%%%%%%%%%%%%%%%%
\section{Experimental setup and procedure} 
\label{sec2}
The experiments have been performed in the rotating tank of the TurLab facility
in Turin. The tank has a diameter of $5 m$ and it rotates anticlockwise with
periods that range from $90$ to $3$ seconds. In the experiment the period of
rotation was set to $T=17.6 s$, corresponding to an angular velocity
$\Omega=2\pi/T = 0.357 rad/s$.

The tank has been filled with fresh water at four different heights 
$H = (10,16,24,32) cm$. Water is seeded with Polyamide particles 
(Arkema Orgasol), with density of 1.03 $g/cm^3$ and diameter 
$d=20 \pm 2 \, \mu m$, which are used for
the visualization of the flow using the Particle Image Velocimetry technique.
The particles are illuminated by an horizontal laser sheet, at $6 cm$ above the
floor of the tank, generated by a Quantum Opus solid state diode green laser.  
The images are acquired by a $8$-bits camera Dalsa Falcon 4M60 with
$2352\times1728$ pixels resolution (further details of the tank and of the
acquisition system can be found in~\cite{Ferrero2009}).  The camera is located
$1.43 m$ above the horizontal laser sheet.

\begin{figure}[h]
\includegraphics[width=0.5\columnwidth]{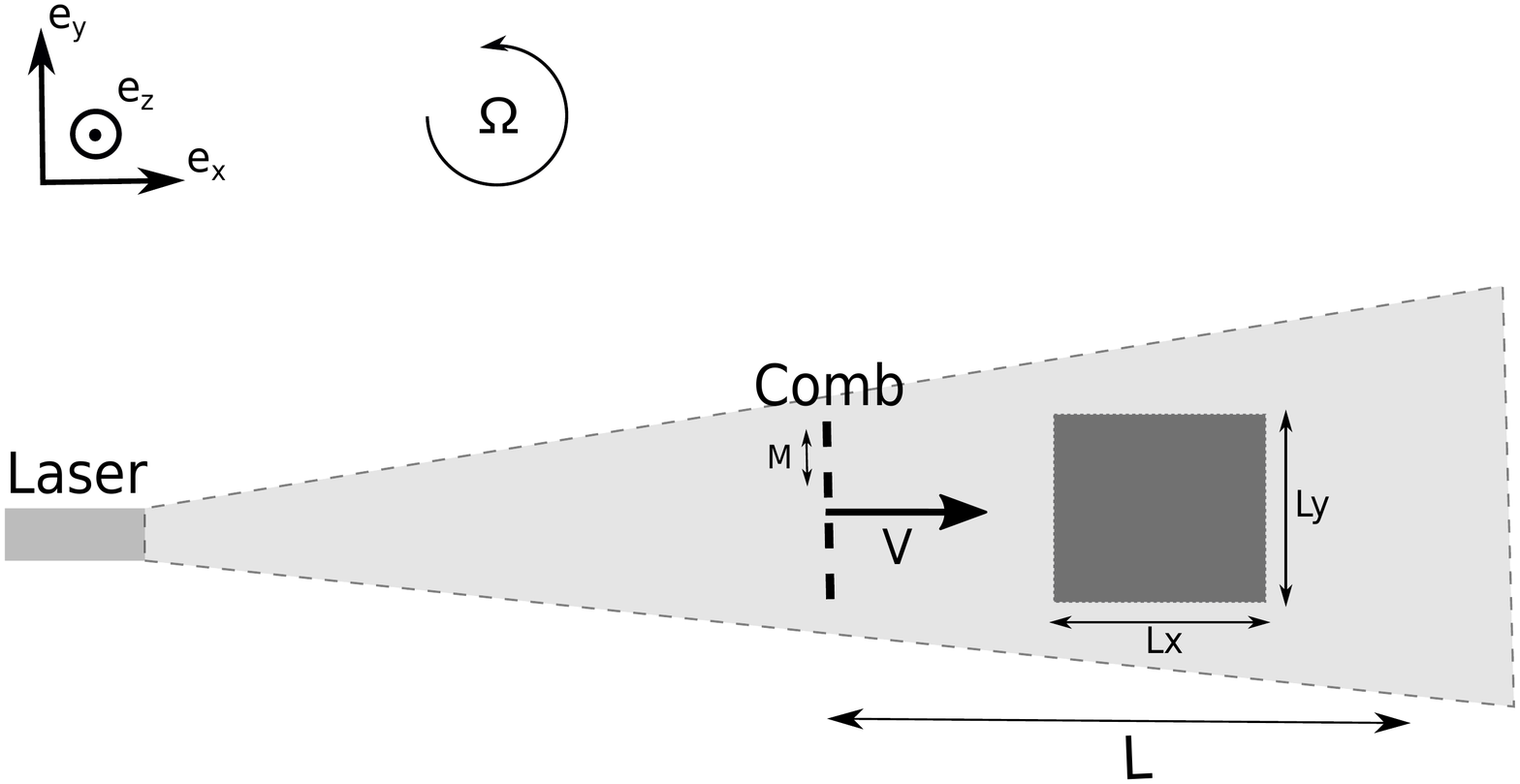}
\includegraphics[width=0.5\columnwidth]{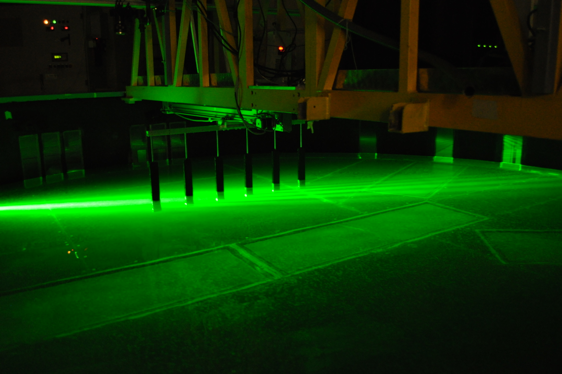}
\caption{Left panel: Schematic of the experimental setup.
Right panel: Photo of the experimental setup}
\label{fig:setup}
\end{figure}

Before the beginning of the experiment, the fluid is set to solid body rotation
by increasing gradually the angular velocity of the tank.
%The focus of the camera is adjusted when the final rotating regime is reached.
Then, trails of vortices are generated by the horizontal motion of a comb,
which is mounted on a motorized linear guide.
The comb is composed by six vertical flat plates of width $a=2.3 cm$
with a mesh size of $M=10 cm$.
It moves with constant velocity $V=18 cm/s$ over a distance $L=90 cm$.
In order to avoid the formation of waves, the velocity of the comb is smoothly
reduced to zero close to the extremities of the guide, before inverting the direction of motion.
The comb Reynolds and Rossby numbers, defined in terms
of the comb velocity $V$ and and the mesh spacing $M$ as in~\cite{Moisy2011}, 
are $Re_c = VM/\nu = 1.8 \times 10^4$ and $Ro_c = V/(2\Omega M) = 2.5$.
A schematic of the experimental setup is represented in Figure~\ref{fig:setup}.
%%%%%%%%%%%%%%%%%%%%%%
%%%%%%%%%%%%%%%%%%%%%%

After 10 minutes of initial forcing, the comb is stopped and the decay of the flow
is recorded for 1 minute with an acquisition rate of $60Hz$. 
The forcing is resumed for a duration of 2 minutes and then stopped before the next recording.
The procedure is repeated 15 times for each height $H$ of the fluid layer.

The velocity fields are obtained by standard PIV analysis, using the
Open Source Particle Image Velocimetry software (OpenPIV, for more detail see~\cite{Taylor2010})
with an interrogation window of 32x32 pixels size and an overlap of 16 pixels.
The resulting velocity fields cover a rectangular area of size $L_x = 28 cm$ and $L_y = 20.5 cm$
and are defined on a grid of $116 \times 85$ points
with a uniform spatial resolution of $\Delta x = \Delta y = 0.241 cm$.
We reconstructed the velocity fields with a sampling rate of $0.1 s$,
skipping an initial time of $0.5 s$ from the last passage of the comb
to avoid the disturbances of the free surface.  

The measured velocity fields 
are the superposition of the turbulent fields ${\bm u}({\bm x},t)$
and a uniform velocity ${\bm U}(t)$ which is due to
the large-scale circulation induced by the comb
and the inertial waves.
The inertial waves manifests in the time series of $U_x(t)$ and $U_y(t)$
as oscillations with period which is half of the rotation period of the tank
$T_{IW} = T/2$ and a phase shift of $\pi/2$ between $U_x$ and $U_y$. 
Before proceeding to the analysis of the data we have subtracted the uniform velocity ${\bm U}(t)$
(as in \cite{Moisy2011}).

The time-series presented in section~\ref{sec4}
are first averaged at fixed time over the ensemble of 15 experiments at given height
and further time-averaged over a window of $1 s$.
The experimental data have been nondimensionalized using
the comb scale $M$,
the rms horizontal velocity at initial time
$u_0 = \langle(u_x^2+u_y^2)/2 \rangle^{1/2}(t=0)$ 
and the timescale $T_0 = M/u_0$.

%%%%%%%%%%%%%%%%%%%%%%%%%%%%%%%%%%%%%%%%%%%%%%%%%%%%%%%%%%%%%%%%%%%%%%%%%%%%
\section{Numerical simulations}
\label{sec3}

Besides the experiments, we also performed a series of direct numerical simulations (DNS)
of an incompressible velocity field in a rotating domain with variable height. 
The dynamics of the velocity field ${\bm u}({\bm x},t)$
is described by the rotating Navier-Stokes equation:
\begin{equation}
\partial_t {\bm u} + {\bm u} \cdot {\bm \nabla} {\bm u} 
+ 2 {\boldsymbol \Omega} \times {\bm u}
= - \frac{{\bm \nabla} p}{\rho} + \nu \nabla^2 {\bm u}
\label{eq1}
\end{equation}
where ${\bf \Omega}=(0,0,\Omega)$ is the angular velocity of the reference frame,
$\rho$ is the uniform density of the fluid, $\nu$ is the kinematic viscosity,
and the pressure $p$ is determined by the condition ${\bf \nabla}\cdot{\bf u}=0$. 

We perform the DNS by means of a standard $2/3$-dealiased, pseudospectral code
with second-order Runge-Kutta integration scheme. 
The velocity field is defined on a triply-periodic domain
with fixed horizontal sizes $L_x=L_y = 2\pi$
and variable height $H = (1/4,1/2,1) \times 2 \pi$. 
It is discretized on a uniform grid at resolution $N_{x}=N_{y}=(H/L_x) N_{z} =512$.
For each height $H$ we consider two values of the angular velocity $\Omega = (1,2)$.
The viscosity is set to $\nu=10^{-3}$. 

At time $t=0$, the velocity field is initialized
as the superposition of a large-scale two-dimensional, two-component (2D2C) flow,
and a small three-dimensional, three-component (3D3C) perturbation.
The 2D2C large-scale flow mimics the 2D vortices generated by the comb in the experiment.
  Nonetheless, it is worth to notice that the initial flows in the DNS and experiments are not identical.
  In the DNS the small 3D perturbation requires some time to develop the 3D turbulent flow. 
  In the experiments, 3D turbulence is already present in the initial flow as a result of the previous passeges of the comb.   
The velocities $u_x$ and $u_y$ of the 2D2C flow are defined in Fourier space as the sum of
random Gaussian horizontal modes $(k_x,k_y,k_z=0)$ with $k_h = (k_x^2+k_y^2)^{1/2}$ in the range $4 < k_h < 6$.
The 3D3D perturbation is defined in the Fourier space as the sum of
random Gaussian modes in the shell $2 < |{\bm k}|< 8$.
The amplitude of the perturbation field is $5 \times 10^{-4}$ smaller than the 2D2C flow.

For each height $H$ we performed 10 simulations with different initial random flow,
keeping constant the kinetic energies of the base flow and of the perturbation.
The time-series presented in section~\ref{sec4} are obtained from the ensemble average
at fixed time of the data obtained in the 10 simulations with given $H$. 
The data of the DNS are nondimensionalized using
the scale $L_0 = 2 \pi /4$, corresponding to the largest wave-length of the initial flow,  
the rms horizontal velocity at initial time $u_0 = \langle(u_x^2+u_y^2)/2 \rangle^{1/2}(t=0)$,
($u_0 = 0.93$ for all $(H,\Omega)$) and the timescale $T_0 = L_0/u_0$.

%%%%%%%%%%%%%%%%%%%%%%%%%%%%%%%%%%%%%%%%%%%%%%%%%%%%%%%%%%%%%%%%%%%%%%%%%%%%
\section{Experimental and numerical results}
\label{sec4}
\begin{figure}[th]
\includegraphics[width=0.5\columnwidth]{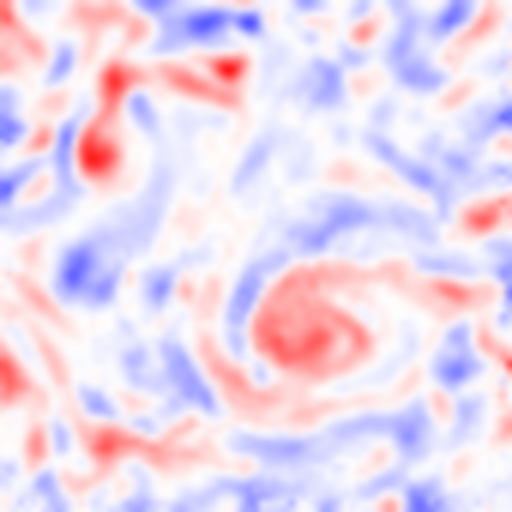}
\includegraphics[width=0.5\columnwidth]{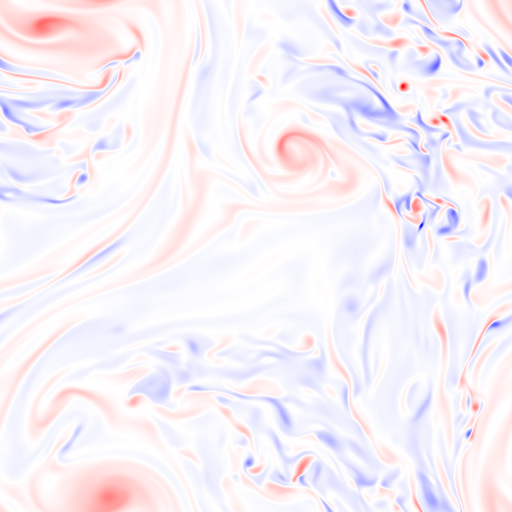}
\caption{
  Vertical vorticity field in the experiments
  with $H=32 cm$ at time $t = 1.8 T_0$ (left panel)
  and in the DNS
  with $H = \pi/2$ and $\Omega=1$ at time $t = 7.1 T_0$
  (right panel).
  Cyclonic vortices are represented in red.  
}
\label{fig:vorticity}
\end{figure}

In Figure~\ref{fig:vorticity} we show two examples of the typical
vorticity fields obtained in the experiments and in the DNS.
More precisely, the left panel shows a square portion (with size $L_y \times L_y$) 
of the vertical vorticity field $\omega_z = \partial_x u_y - \partial_y u_x$
at time $t = 1.8 T_0$ in the experiments at $H=32 cm$,
while the right panel shows a section at $z=0$
of the vertical vorticity field $\omega_z(x,y,z=0)$
at time $t = 7.1 T_0$ in the simulations at $H = \pi/2$ and $\Omega=1$.
In both the experiments and the DNS it is clearly visible the presence 
of large-scale cyclonic vortices (represented in red).
%%%%%%%%%%%%%%%%%%%%%%
%%%%%%%%%%%%%%%%%%%%%%

\begin{figure}[h]
  \includegraphics[width=0.5\columnwidth]{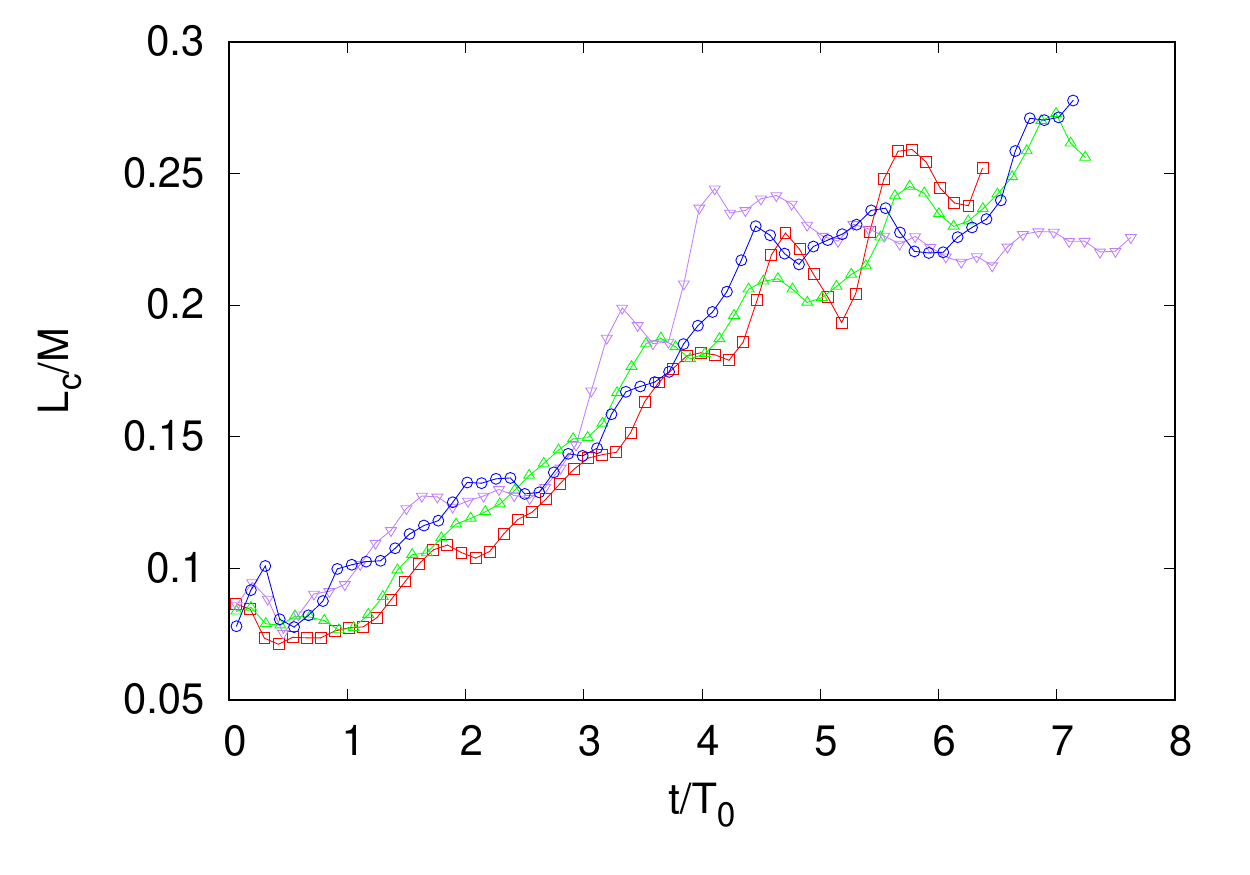}
  \includegraphics[width=0.5\columnwidth]{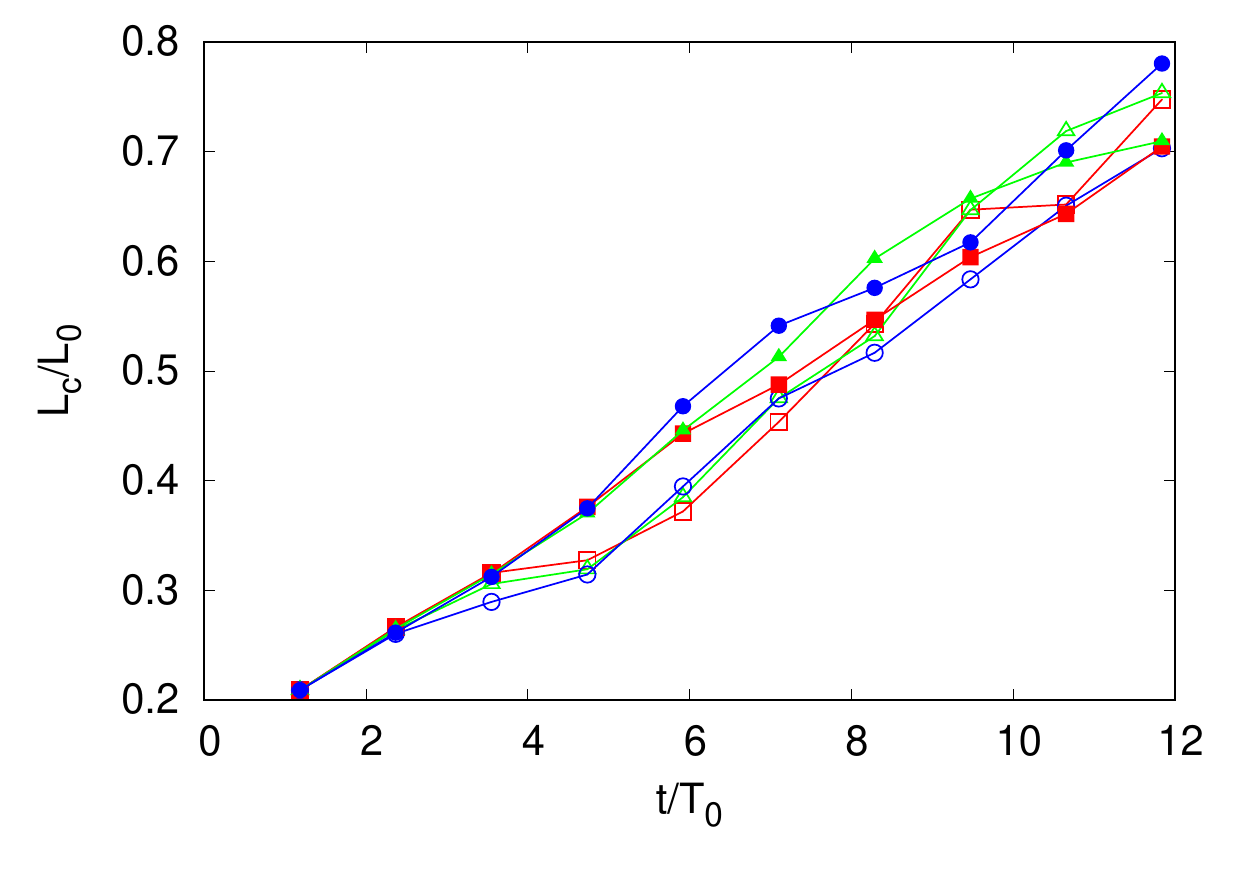}  
  \caption{Left panel: Velocity correlation length $L_c$
in the experiments at $\Omega=0.357 rad/s$ with 
$H=10 cm$ (red squares),
$H=16 cm$ (green triangles),
$H=24 cm$ (purple down-pointing triangles) and
$H=32 cm$ (blue circles).
Right panel: Velocity correlation length $L_c$
in the DNS with
$H=\pi/2 $ (red squares),
$H=\pi$ (green triangles) and
$H=2\pi$ (blue circles)
at angular velocity $\Omega=1$ (empty symbols)
and $\Omega =2$ (filled symbols).
  }
\label{fig:lcorr}
\end{figure}

The formation of these structures during the decay of the
rotating flow causes an increase of the horizontal correlation scale.
In order to quantify this effect we first compute
the longitudinal correlation function of horizontal velocity 
$C(r,t) = \langle u_\alpha({\bm x},t)u_\alpha({\bm x}+r{\bm e}_\alpha,t)\rangle
/\langle  u_\alpha({\bm x},t)^2 \rangle$
with $\alpha = (x,y)$.   
Then we define the correlation length $L_c(t)$
as the scale at which $C(L_c) = 0.8$.
The time evolution of $L_c$ is shown in Figure~\ref{fig:lcorr}.
In both the experiment and the DNS we observe a weak dependence of $L_c$ on $H$.
The scale $L_c$ increases almost linearly in time for $t > T_0$.
Previous studies have reported a different scaling $L_c(t) \simeq t^\beta$
with exponent $\beta$ in the range $(0.2,0.4)$~\cite{jacquin1990homogeneous,Moisy2011}.
We note that the growth of $L_c$ is faster in the DNS than in the experiments:
In the DNS the average growth rate of $L_c$ is $L_c/L_0 \simeq 0.05 t/T_0$, while
in the experiments it is $L_c/M \simeq 0.03 t/T_0$.
This effect could be caused by the 2D2C initial condition in the DNS,
which induce a 2D dynamics characherized by stronger large-scale energy transfer.
%We note the value of the exponent might be affected by the initial offset
%of the scale $L_c$ and by the choice of the initial time. 
%%%%%%%%%%%%%%%%%%%%%
%%%%%%%%%%%%%%%%%%%%%

The growth of the correlation scale influences the time evolution
of the Reynolds and Rossby numbers, defined as
\begin{equation}
  Re(t) = \frac{u_h L_c}{\nu}\;,\;\;
  Ro(t) = \frac{u_h}{2\Omega L_c}\;,
  \label{eq:rero}
\end{equation}
where $u_h$ is the rms horizontal velocity
$u_h = (\langle (u_x^2 + u_y^2)/2\rangle^{1/2})$. 
%We remind that in the experiment ${\bm u}$
%is the fluctuating part of the velocity field obtained by
%subtracting the uniform velocity due to the large-scale circulation
%and the gravity waves.
As shown in Figure~{\ref{fig:re} (left panel), 
in the experiments, after an initial rapid decay at $t < T_0$,
the Reynolds number remains approximatively constant with some fluctuations
(similarly to what observed in~\cite{Moisy2011}). 
Conversely, in the DNS we observe an almost linear
increase of $Re(t)$, which indicates that the growth of $L_c(t)$
overwhelms the decay of the velocities.
  We argue that the difference between the behavior of two systems
  could be ascribed to their different boudary conditions.
  In the experiments the bottom friction (which is absent in the DNS)
  causes a faster decay of the velocities, resulting in a different temporal evolution of $Re$.
The dependence of $Re(t)$ on $H$ is unclear: in the experiments
with the thinner layer ($H=10 cm$) the values of $Re$ are on average smaller
than those measured with the thickest layer ($H=32 cm$),
but we observe the opposite behavior in the DNS with $\Omega=1$.
  
%%%%%%%%%%%%%%%%%%%%%
\begin{figure}[h]
  \includegraphics[width=0.5\columnwidth]{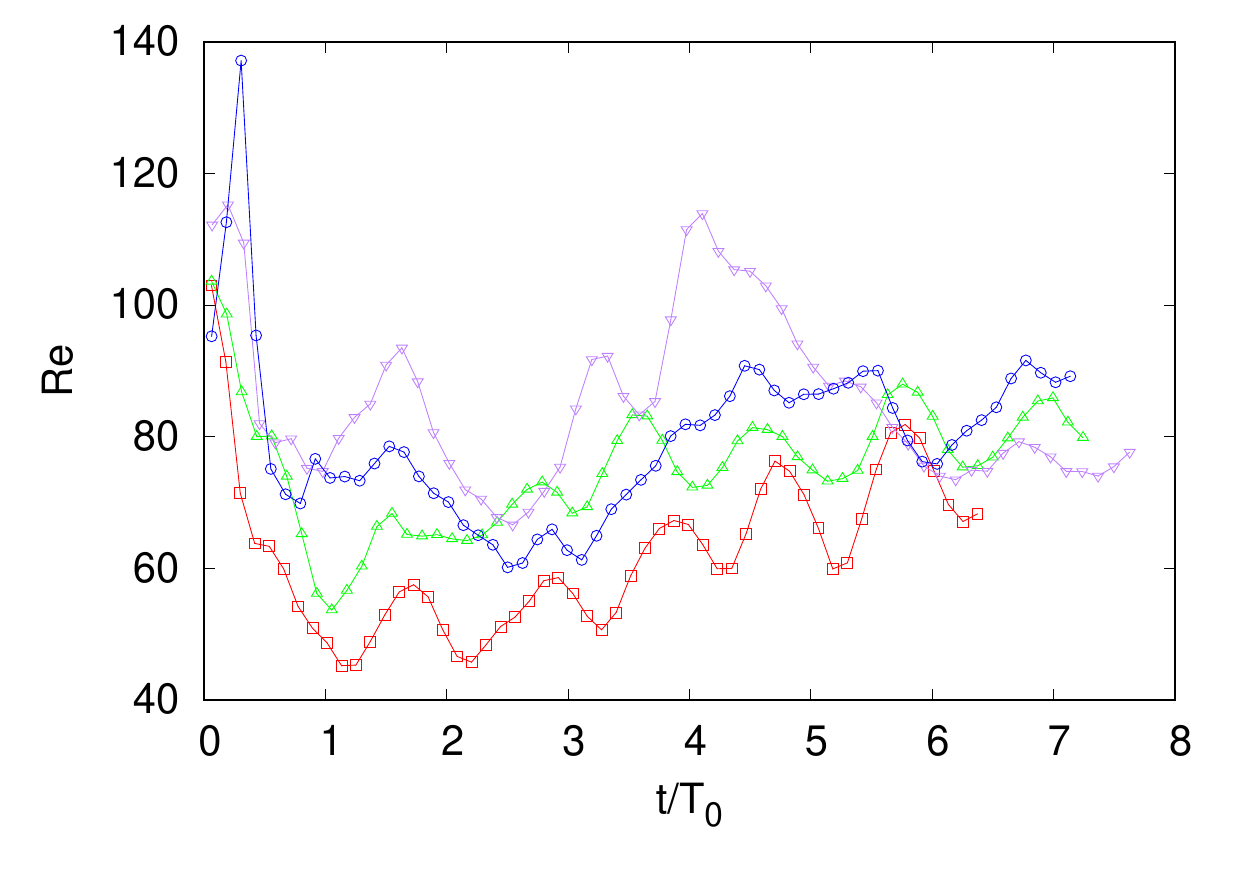}
  \includegraphics[width=0.5\columnwidth]{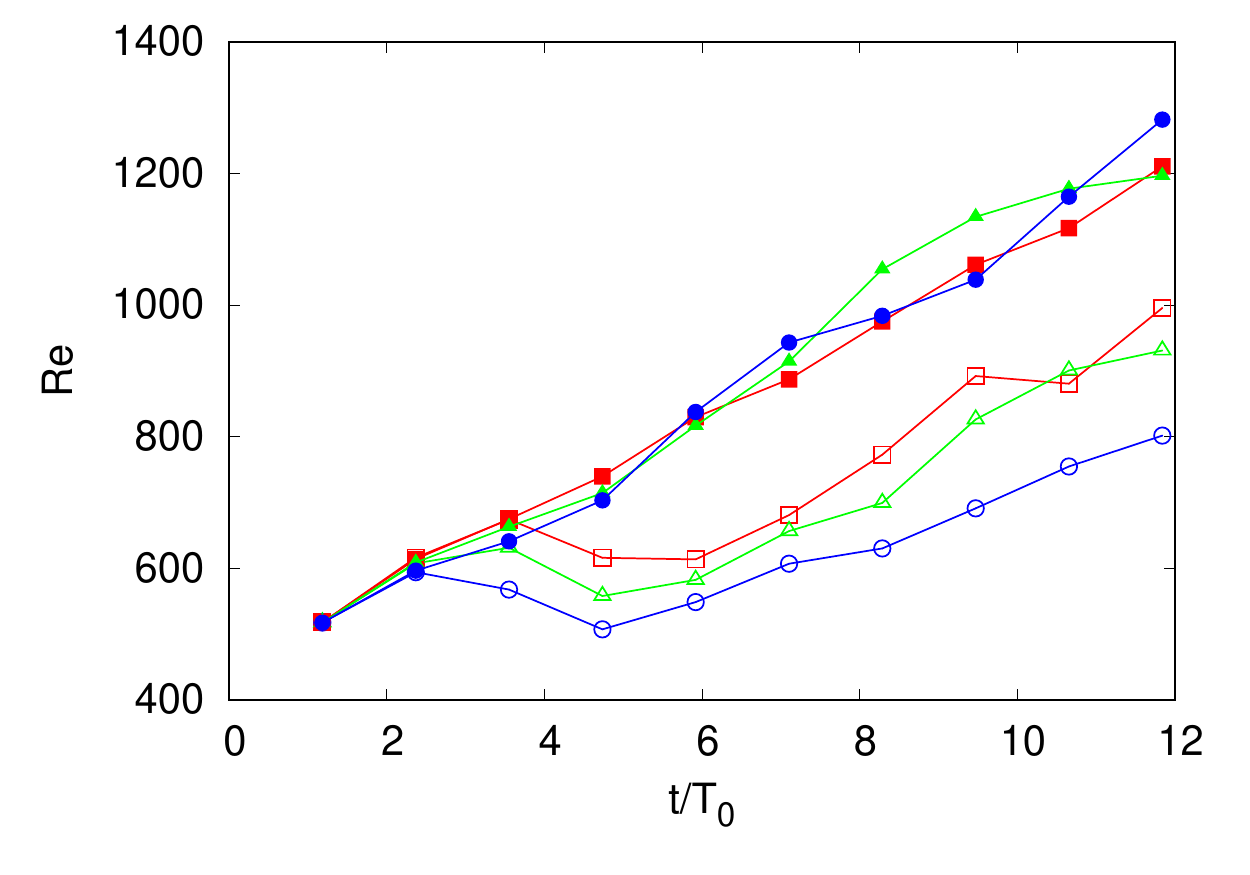}  
  \caption{Reynolds number $Re = L_c u_{rms}/\nu$ 
    in the experiments (left panel) 
     and in the DNS (right panel)
    at angular velocity $\Omega=1$ (empty symbols)
    and $\Omega =2$ (filled symbols).
    Symbols as in Figure~\ref{fig:lcorr}.
  }
\label{fig:re}
\end{figure}
%%%%%%%%%%%%%%%%%%%%%

The Rossby number decreases in time both in the experiments
and in the numerics (see Figure~\ref{fig:ro})
and it is almost independent on $H$.
At long times $t > T$ we observe a scaling regime $Ro(t) \simeq t^{-1}$. 
This scaling has been previously reported in~\cite{Moisy2011}.
The decay of the Rossby number indicates that the Coriolis force prevails
over the inertial forces at long times.
Therefore the effects of rotation are expected to
becomes more pronounced as the system evolves.
%%%%%%%%%%%%%%%%%%%%%
\begin{figure}[h]
\includegraphics[width=0.5\columnwidth]{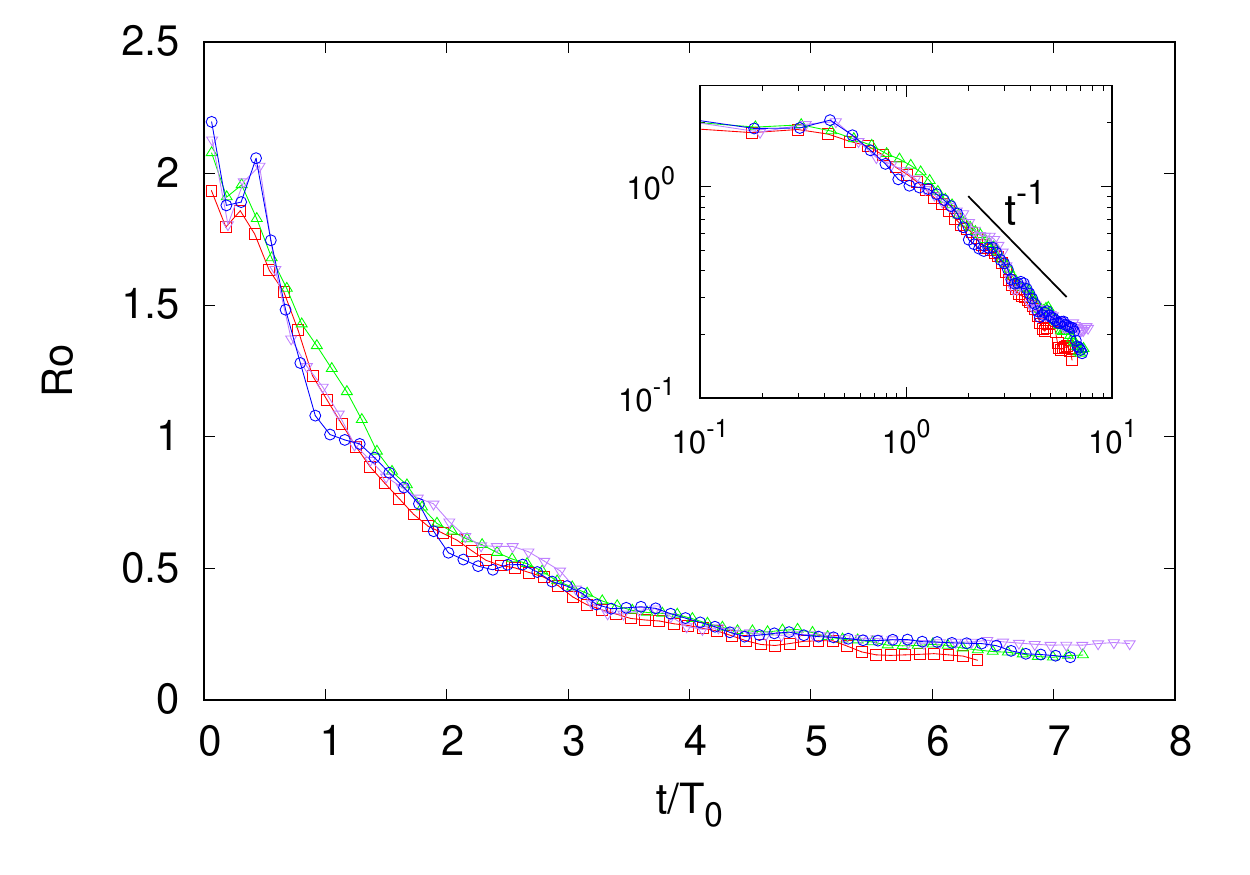}
\includegraphics[width=0.5\columnwidth]{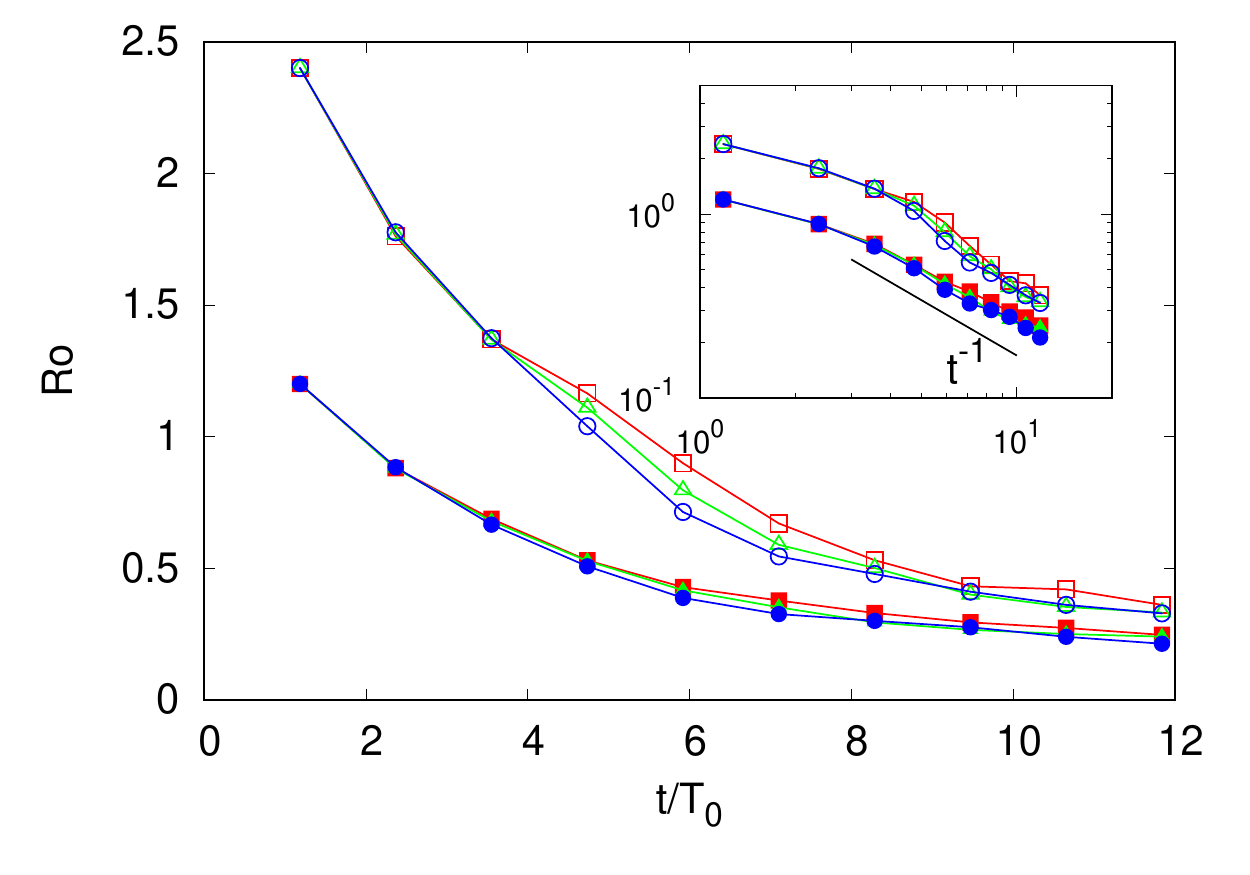}
  \caption{Rossby number $Ro = u_{rms}/2\Omega L_c$ 
    in the experiments (left panel)
    and in the DNS (right panel)
    at angular velocity $\Omega=1$ (empty symbols)
    and $\Omega =2$ (filled symbols).
    Symbols as in Figure~\ref{fig:lcorr}.
  }
\label{fig:ro}
\end{figure}
%%%%%%%%%%%%%%%%%%%%%

In Figure~\ref{fig:spettri} we compare the energy spectra $E_h(k)$
of the horizontal velocity fields $u_x,u_y$
in the experiments at $t = 1.8 T_0$ (left panel) 
and in the DNS at time $t = 7.1 T_0$ (right panel).
In the experiments we observe a power-law spectrum $E_h(k) \simeq k^{-2}$
in the wavenumber range $6 < k M < 20$.
A similar spectral slope is observed also in the DNS at $\Omega=1$
in the range  $10 < k L_0 < 30$,
while the simulations with $\Omega=2$ have steeper spectra. 
In both the experiments and DNS the spectra are almost independent
on the heights $H$.
In the spectra of the DNS it is possible to observe a beginning of  
accumulation of energy in the lowest accessible mode.
The spectral condensation is clearly visible in the spectra at late times of the DNS (not shown). 
Conversely, this phenomenon is not observed in the experiments
because of the large scale separation between the diameter of the tank and
the typical size of the vortices produced by the comb. 
%%%%%%%%%%%%%%%%%%%%%
\begin{figure}[h]
\includegraphics[width=0.5\columnwidth]{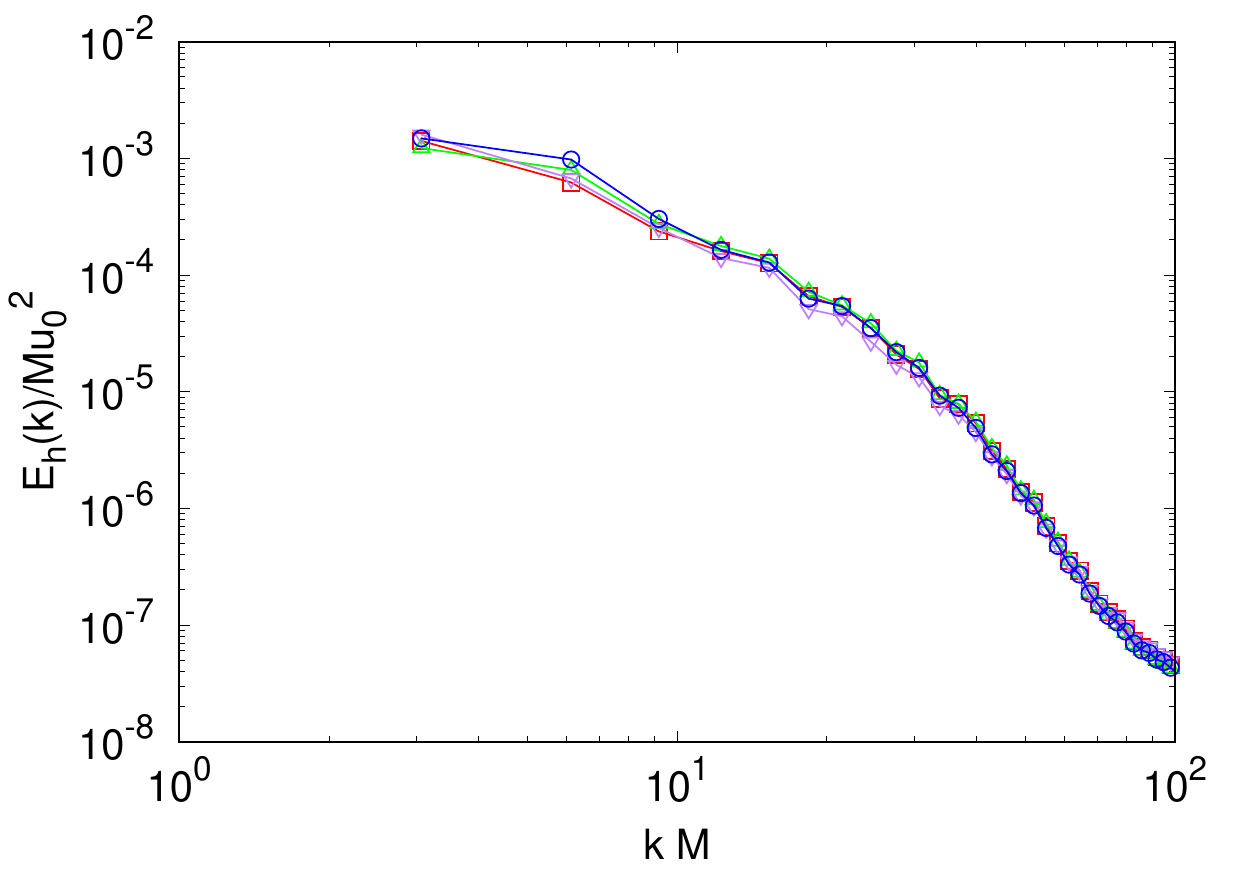}
\includegraphics[width=0.5\columnwidth]{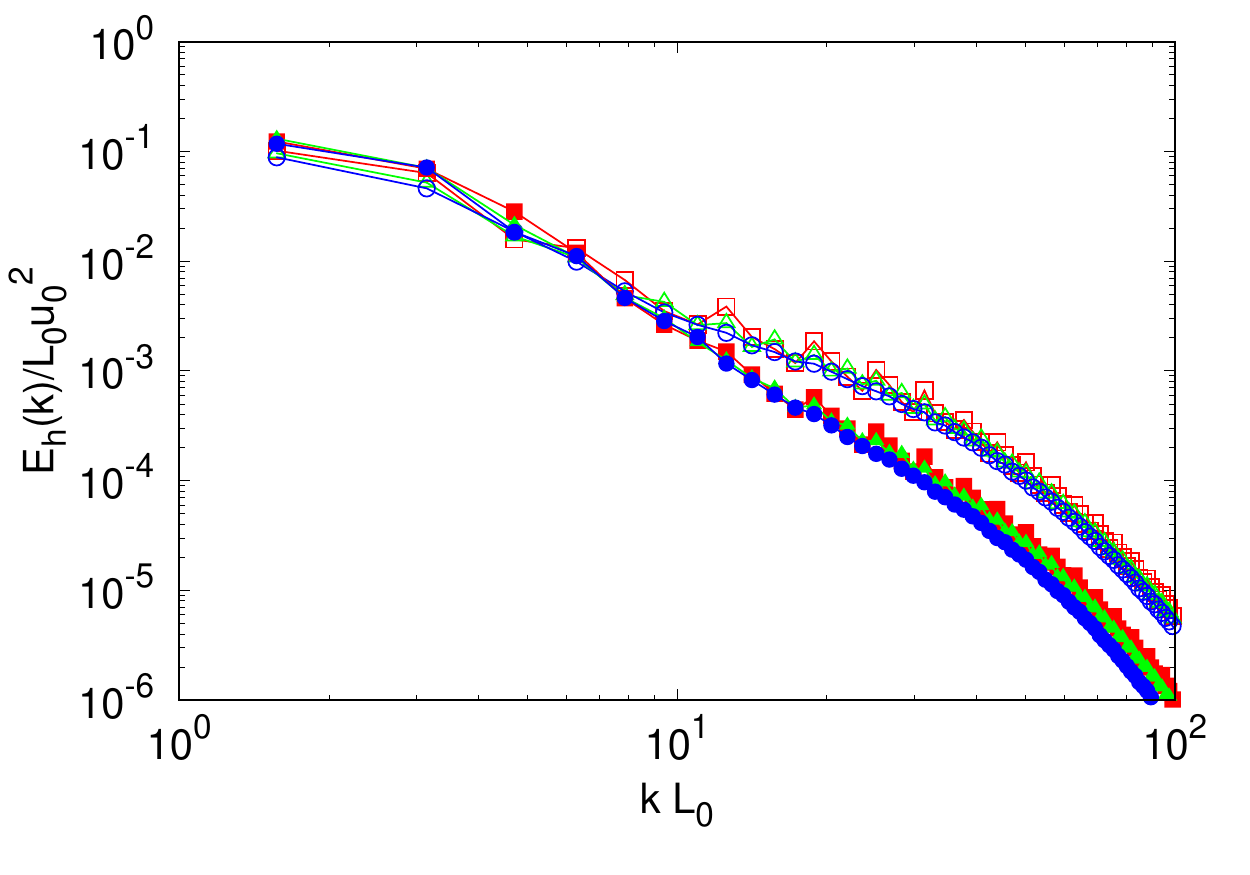}
  \caption{Horizontal energy spectra 
    in the experiments at time $t = 1.8 T_0$ (left panel)
    and in the DNS at (right panel)
    at angular velocity $\Omega=1$ (empty symbols)
    and $\Omega =2$ time $t = 7.1 T_0$ (filled symbols).
    Symbols as in Figure~\ref{fig:lcorr}.
  }
\label{fig:spettri}
\end{figure}
%%%%%%%%%%%%%%%%%%%%%

The results presented so far do not show a strong dependence on the height of the fluid layer.
On the contrary, the effect of varying $H$ is clearly visible in the statistics of the
vertical component of the vorticity $\omega_z = \partial_x u_y - \partial_y u_x$.
The probability distribution functions (PDF) of $\omega_z$ are shown in Figure~\ref{fig:pdf}
for different values of $H$ at a fixed time
$t = 1.8 T_0$ in the experiments and
$t = 7.1 T_0$ in the DNS.
The PDFs corresponding to the large $H$ are characterized by a positive skewness
$S_\omega = \langle \omega_z^3 \rangle/\langle \omega_z^2 \rangle^{3/2}$,
which quantifies the cyclone-anticyclone asymmetry.
Reducing the thickness $H$, the PDFs become more symmetric and the skewness is reduced.
This means that the confinement of the decaying flow in a thin layer
weakens the cyclone-anticyclone asymmetry at fixed time. 
This is in qualitative agreement with previous numerical results in forced 
stationary conditions \cite{Deusebio2014}.
%%%%%%%%%%%%%%%%%%%%%
\begin{figure}[h]
\includegraphics[width=0.5\columnwidth]{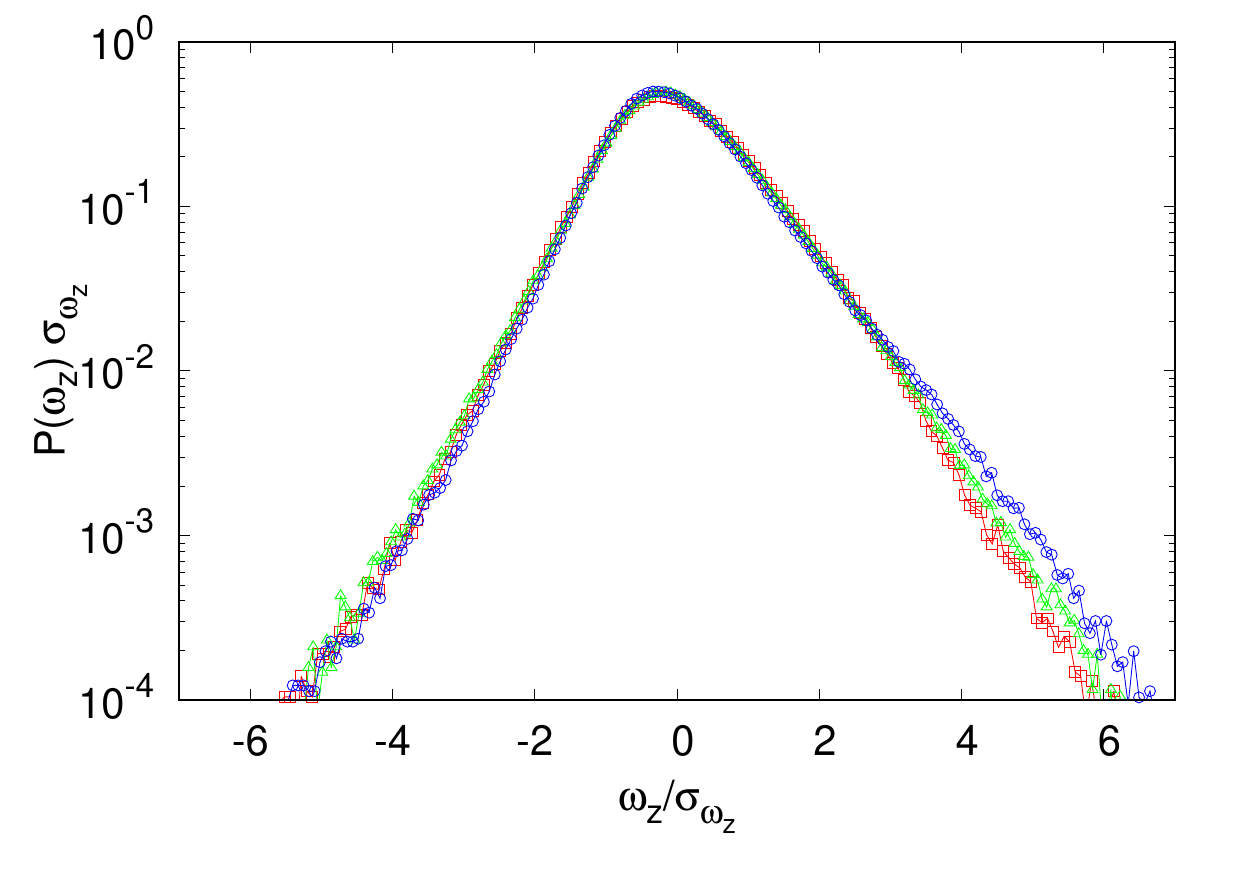}
\includegraphics[width=0.5\columnwidth]{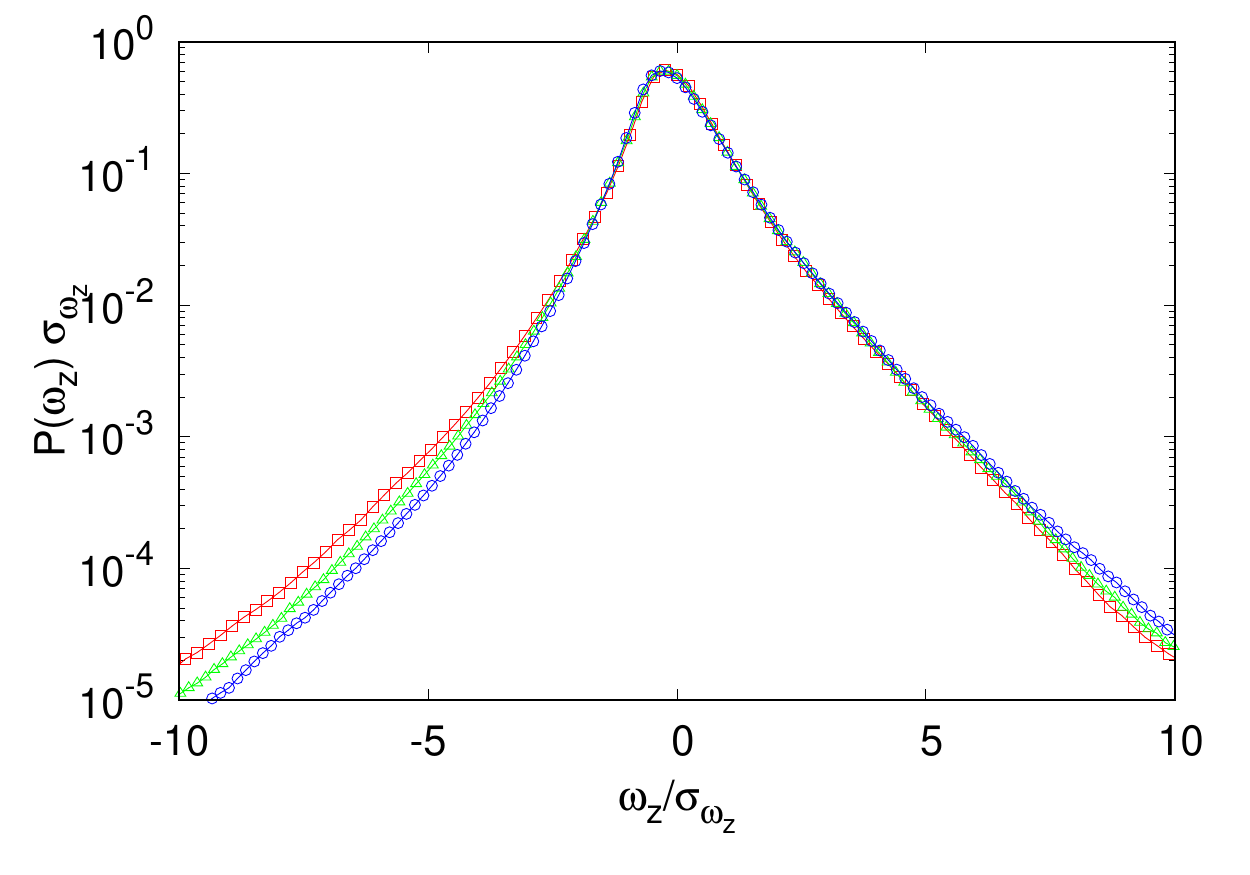}
  \caption{PDFs of the vertical vorticity ${\omega_z}$ 
    in the experiments at time $t = 1.8 T_0$ (left panel)
    and in the DNS at time $t = 7.1 T_0$ at angular velocity $\Omega=1$ (right panel).
    Symbols as in Figure~\ref{fig:lcorr}.
  }
\label{fig:pdf}
\end{figure}
%%%%%%%%%%%%%%%%%%%%%

%%%%%%%%%%%%%%%%%%%%%
\begin{figure}[h]
\includegraphics[width=0.5\columnwidth]{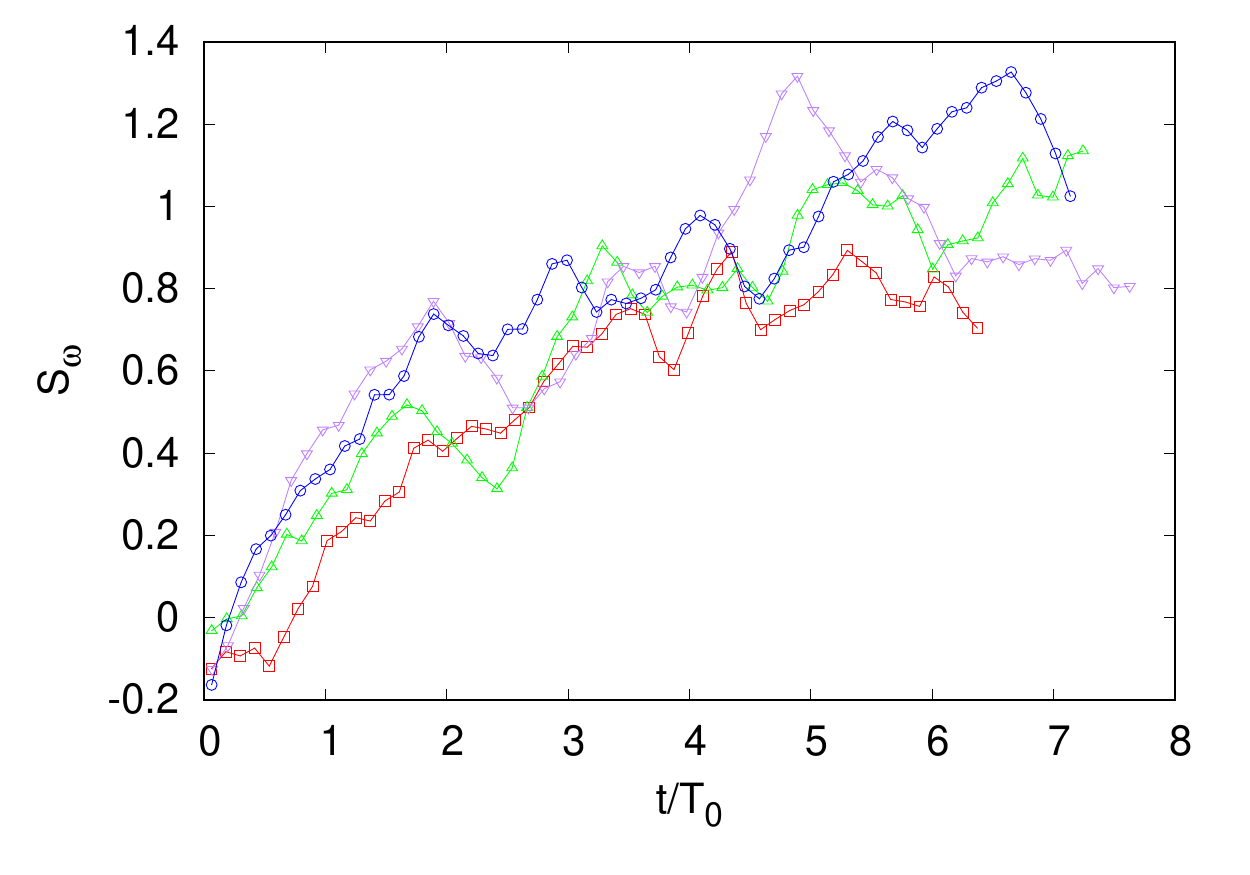}
\includegraphics[width=0.5\columnwidth]{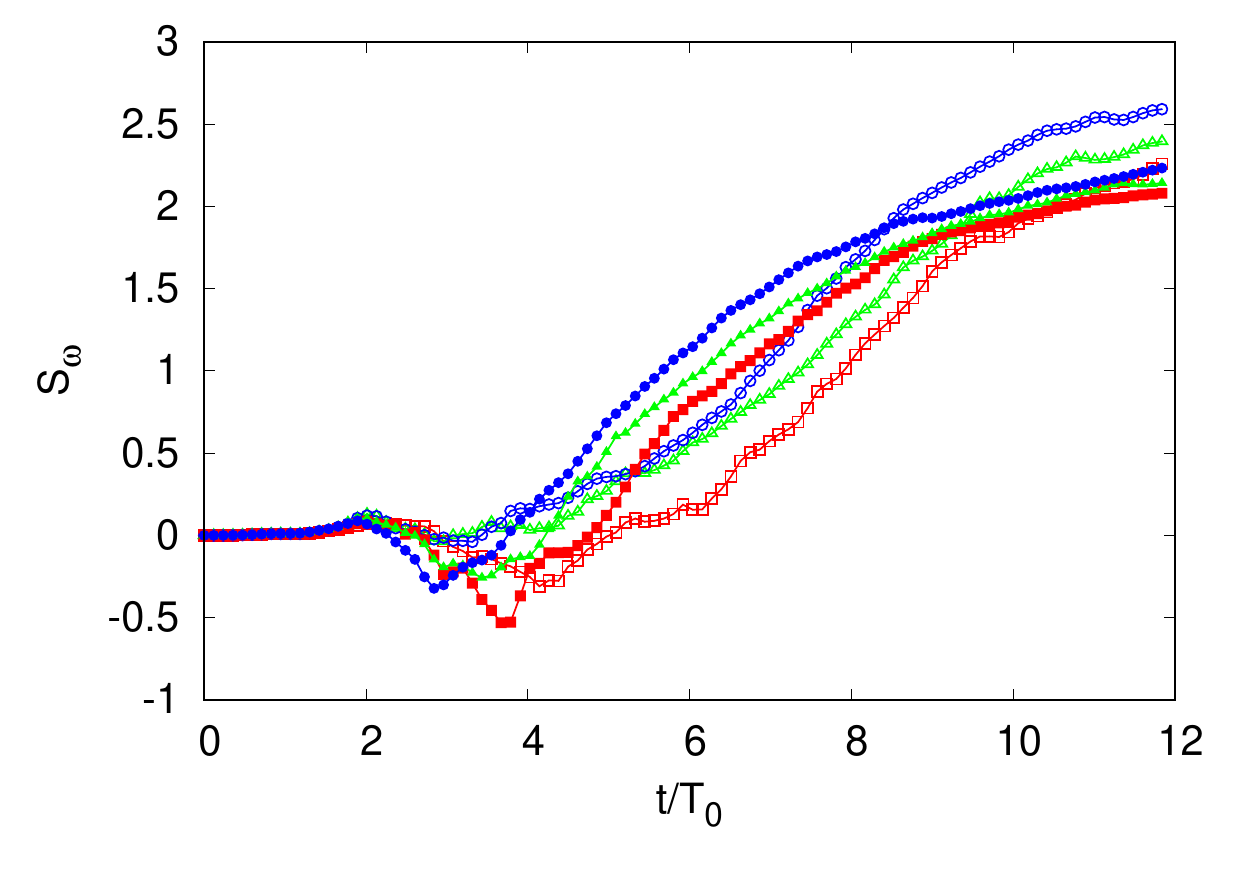}
  \caption{Skewness of the vertical vorticity ${\omega_z}$ 
    in the experiments (left panel)
    and in the DNS (right panel)
    at angular velocity $\Omega=1$ (empty symbols)
    and $\Omega =2$ (filled symbols).
    Symbols as in Figure~\ref{fig:lcorr}.
  }
\label{fig:skew}
\end{figure}
%%%%%%%%%%%%%%%%%%%%%
Because of the decay of the Rossby number,
it is expected that the cyclone-anticyclone asymmetry increases with time. 
Previous studies\cite{Moisy2011,Naso2015} reported a power-law growth of the
skewness $S_\omega \simeq t^{\gamma}$ with $\gamma \approx 0.70 \pm 0.05$. 
Here we are interested to investigate how the height $H$ of the fluid layer
influences the growth of $S_\omega$.
The temporal evolution of $S_\omega$ is shown in Figure~\ref{fig:skew}.
In all the simulations and experiments, after an initial transient we observe the development
of a positive skewness, which indicates the prevalence of cyclones over anticyclones.
In the DNS, we find that the regime of positive skewness is systematically preceded by a transient
in which $S_\omega$ is negative. We are not aware of previous observations of this phenomenon.
After the negative transient, the skewness in the DNS grows as $S_\omega(t) \sim (t-t_*)^{0.80\pm 0.05}$
(not shown), being $t_*$  the time at which $S_\omega$ changes sign from negative to positive.
The value of the exponent is in agreement with the results reported in \cite{Naso2015}.

The series of $S_\omega(t)$ obtained in the numerics display a clear dependence on $H$.
Smaller $H$ correspond to smaller values of $S_\omega$ at fixed short time.
A similar dependence on $H$ is observed also in the experimental series, even if they
are more noisy.
After the initial growth, the skewness saturates to almost constant values at late times
both in the DNS (for $t > 8 T_0$) and experiments (for $t> 4 T_0$).
In the numerical series with $\Omega=2$ the asymptotic value
of the skewness has not a clear dependence on $H$.
In the experiments with $H=24 cm$ we observe a decay of $S_\omega$ at $t > 5 T_0$.
It is tempting to interpret this as the beginning of the long time decay of the vorticity skewness
which has been reported in previous studies (e.g.\cite{Moisy2011}). 
Nonetheless, even after averaging over 15 independent experiments,
our data displays strong temporal fluctuations which do not allow
to make accurate statements concerning the late stage of the evolution of the skweness.
The inspection of the numerical series in Figure~\ref{fig:skew} suggests that,
while the cyclone-anticyclone asymmetry develops for all the cases with different $H$ considered here,
reaching similar values of $S_\omega$ at the end of the simulations, 
the main effect of the confinement of the flow in a thin layer is to retard its development.
To test this idea, in Figure~\ref{fig:skew_resc} we plot the series of $S_\omega(t)$
by rescaling the times with height-dependent time scales $T_H$.
The values of $T_H$ have been determined by least square method,
minimizing the differences between $S_\omega(t/T_H)$ at given $H$
with respect to the case with the largest $H=H_{max}$
($H_{max} = 32 cm$ in the experiments and $H_{max} = 2\pi$ in the DNS),
and fixing $T_{H_{max}} = T_0$. 
The collapse of the series is reasonably good,
and the rescaling times $T_H$ become larger as the thickness $H$ is reduced.
We note that the values of $T_H/T_0$ are identical
in the DNS with $\Omega=1$ and $\Omega=2$.
This shows that the confinement in a thin layer slows down
the development of the cyclone-anticyclone asymmetry.
It is interesting to note that this effect is qualitatively
similar in the experiment and in the DNS, in spite of the
differences bewteeen the two systems highlighted
in the introduction and observed in the temporal evolution
of the correlation scale (Fig.~\ref{fig:lcorr})
the Reynolds mumber (Fig.~\ref{fig:re})
and the spectra (Fig.~\ref{fig:spettri}). 

%%%%%%%%%%%%%%%%%%%%%
\begin{figure}[h]
\includegraphics[width=0.5\columnwidth]{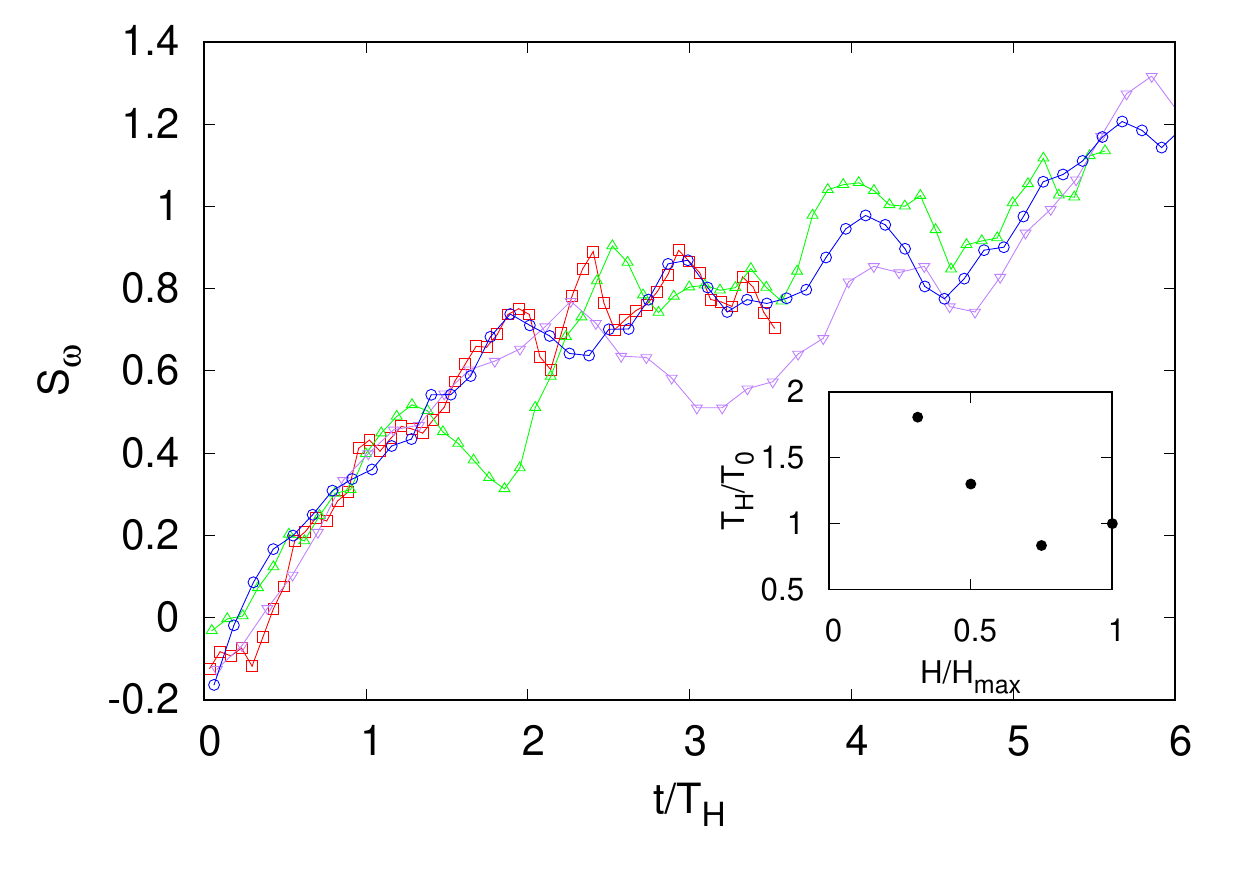}
\includegraphics[width=0.5\columnwidth]{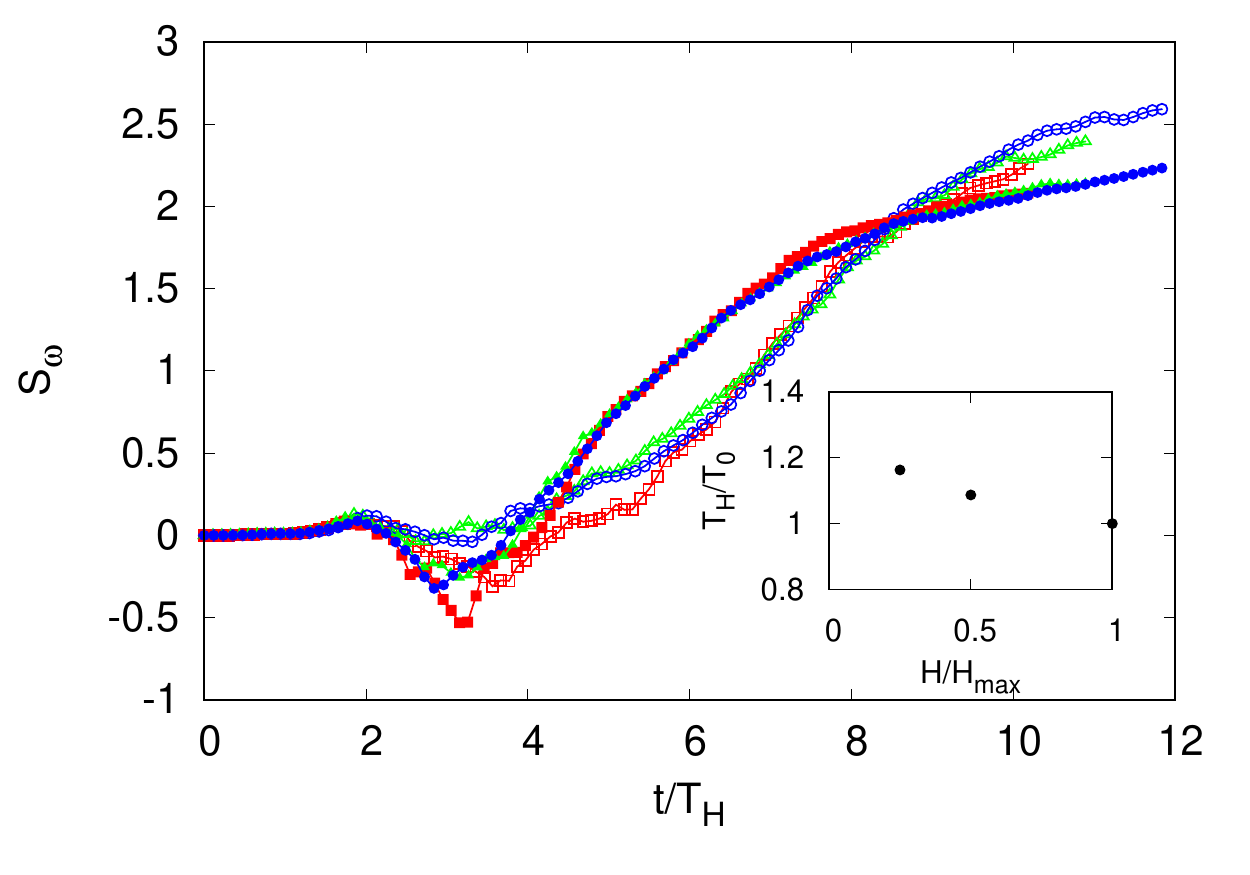}
  \caption{Skewness of the vertical vorticity ${\omega_z}$ 
    in the experiments (left panel)
    and in the DNS (right panel)
    at angular velocity $\Omega=1$ (empty symbols)
    and $\Omega =2$ (filled symbols).
    Time has been rescaled with $T_H$.
    The values of $T_H$ are shown in the insets. 
    Symbols as in Figure~\ref{fig:lcorr}.
  }
\label{fig:skew_resc}
\end{figure}
%%%%%%%%%%%%%%%%%%%%%

Finally we present a result of the late-stage of the decay in the DNS.
We have continued the DNS up to time $t=24 T_0$.
At that time, the turbulent fluctuations are almost completely disappeared, 
and the velocity field consists of a single cyclonic vortex.
Because in the DNS the mean vorticity is constrained to be zero,
the vortex is surrounded by a sea of negative vorticity.
As one can see in Figure~\ref{fig:pdf_cutoff}, the PDF of the vorticity 
field of this fossil state of turbulence
displays an interesting feature: its negative tail has a sharp cutoff
at $\omega_z = -2 \Omega$.
In other words, at long times the total vorticity computed in the laboratory frame
$\omega_z + 2\Omega$ is always positive.
This result contrasts with the recovery of the symmetry at long times 
which has been observed in ~\cite{Morize2005,Moisy2011}.
As discussed in Ref.~\cite{Moisy2011},
the symmetry is expected to be restored only if the initial state contains
a significant amount of vertical velocity, which is almost absent in our case. 
It would be interesting to investigate more systematically how this phenomenon
is dependent on the properties of the initial velocity field
and on the boundary conditions. 
%%%%%%%%%%%%%%%%%%%%%
\begin{figure}[h]
\includegraphics[width=0.5\columnwidth]{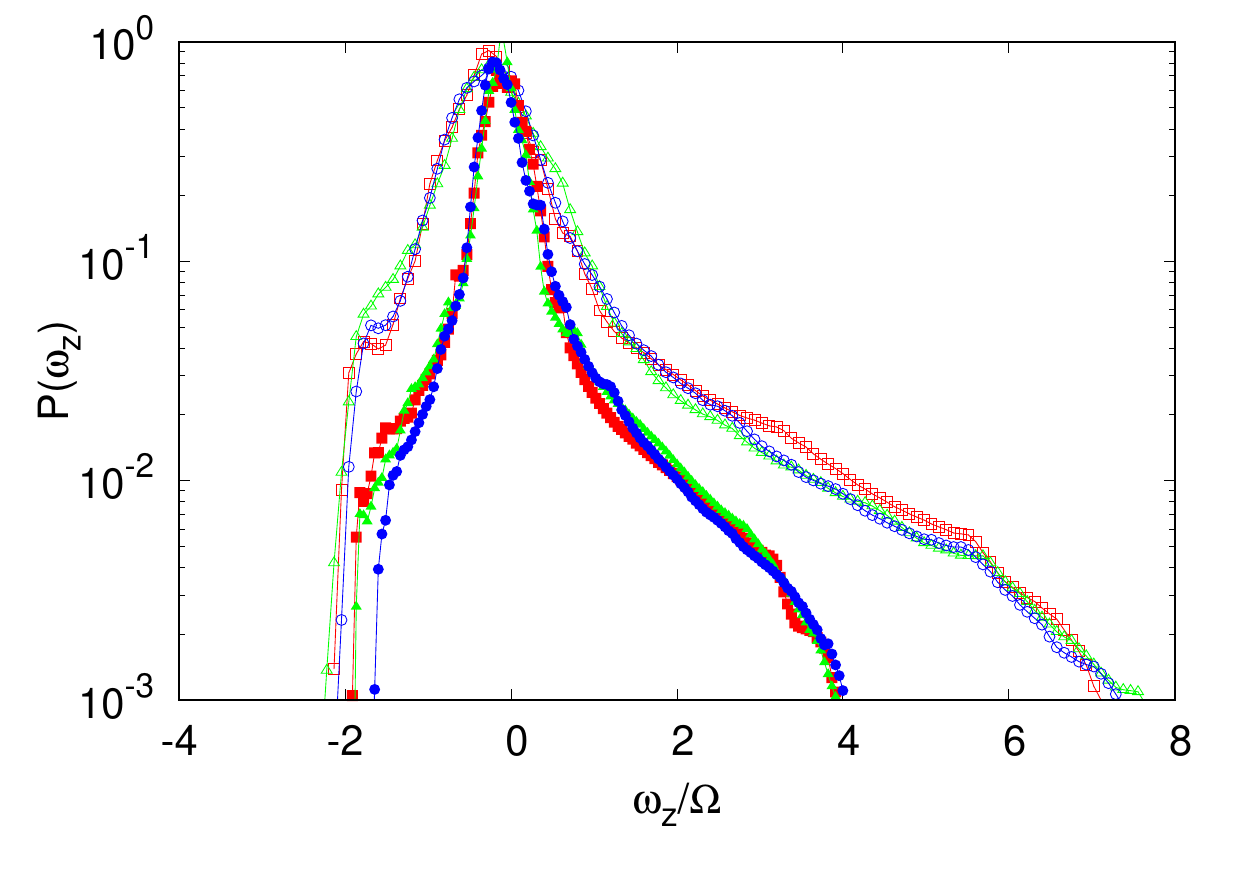}
\caption{PDFs of the vertical vorticity ${\omega_z}$
    in the numerical simulations
    at time $t= 24 T_0$
    at angular velocity $\Omega=1$ (empty symbols)
    and $\Omega =2$ (filled symbols).
   Symbols as in Figure~\ref{fig:lcorr}. 
  }
\label{fig:pdf_cutoff}
\end{figure}
%%%%%%%%%%%%%%%%%%%%%

\section{Conclusion}
\label{sec5}
The main result of our study is that the confinement of a turbulent rotating
flow in a thin layer delays the development of the cyclone-anticyclone 
asymmetry.
This effect is observed both in experiments and in numerical
simulations which have structural differences in the boundary conditions, 
and it is therefore a robust feature of decaying rotating flows,
independent on the presence of bottom friction.
Our findings show that the, although the formation
of the cyclone-anticyclone asymmetry is obseved both with and without vertical confinement,
the height of the fluid layer is a crucial parameter to determine the temporal scale of this phenomenon. 
Further experiments and numerical simulations are needed to better understand how the mechanism
of formation of cyclonic columnar structures are influenced by the vertical confinement.

Our results have important implication for large-scale geophysical flows,
where the height of the fluid layer is typically smaller than the horizontal
scales. The cyclone-anticyclone asymmetry observed in these conditions 
could be much weaker than what expected on the basis of experiments and 
DNS with aspect ratio of order unity.

%%%%%%%%%%%%%%%%%%%%%%%%%%%%%%%%%%%%%%%%%%%%%%%%%%%%%%%%%%%%%%%%
\section*{Acknowledgment(s)}

We acknowledge the financial support of the project "European High performance Infrastructures
in Turbulence" (EUHIT) (Grant agreement ID: 312778) in the frame of the Research
Infrastructures Integrating Activity framework of FP7.
We also acknowledge support by the Departments of Excellence grant (MIUR).

%%%%%%%%%%%%%%%%%%%%%%%%%%%%%%%%%%%%%%%%%%%%%%%%%%%%%%%%%%%%%%%%
\bibliographystyle{tfnlm}
\bibliography{biblio}

\begin{thebibliography}{10}
\providecommand{\url}[1]{\normalfont{#1}}
\providecommand{\urlprefix}{Available from: }

\bibitem{Hopfinger1982}
Hopfinger~EJ, Browand~FK, Gagne~Y. Turbulence and waves in a rotating tank.
  Journal of Fluid Mechanics. 1982;\hspace{0pt}125:505--534.

\bibitem{Longhetto2002}
Longhetto~A, Montabone~L, Provenzale~A, et~al. Coherent vortices in rotating
  flows: a laboratory view. Nuovo Cimento della Societa Italiana di Fisica C.
  2002;\hspace{0pt}25(2).

\bibitem{Staplehurst2008}
Staplehurst~PJ, Davidson~PA, Dalziel~SB. Structure formation in homogeneous
  freely decaying rotating turbulence. Journal of Fluid Mechanics.
  2008;\hspace{0pt}598:81--105.

\bibitem{Moisy2011}
Moisy~F, Morize~C, Rabaud~M, et~al. Decay laws, anisotropy and
  cyclone–anticyclone asymmetry in decaying rotating turbulence. Journal of
  Fluid Mechanics. 2011;\hspace{0pt}666:5--35.

\bibitem{Bartello1994}
Bartello~P, M\'etais~O, Lesieur~M. Coherent structures in rotating
  three-dimensional turbulence. Journal of Fluid Mechanics.
  1994;\hspace{0pt}273:1--29.

\bibitem{Yeung1998}
Yeung~P, Zhou~Y. Numerical study of rotating turbulence with external forcing.
  Physics of Fluids. 1998;\hspace{0pt}10(11):2895--2909.

\bibitem{Smith1999}
Smith~LM, Waleffe~F. Transfer of energy to two-dimensional large scales in
  forced, rotating three-dimensional turbulence. Physics of fluids.
  1999;\hspace{0pt}11(6):1608--1622.

\bibitem{Yoshimatsu2011}
Yoshimatsu~K, Midorikawa~M, Kaneda~Y. Columnar eddy formation in freely
  decaying homogeneous rotating turbulence. Journal of fluid mechanics.
  2011;\hspace{0pt}677:154.

\bibitem{Biferale2016}
Biferale~L, Bonaccorso~F, Mazzitelli~IM, et~al. Coherent structures and extreme
  events in rotating multiphase turbulent flows. Physical Review X.
  2016;\hspace{0pt}6(4):041036.

\bibitem{Godeferd2015}
Godeferd~FS, Moisy~F. Structure and dynamics of rotating turbulence: a review
  of recent experimental and numerical results. Applied Mechanics Reviews.
  2015;\hspace{0pt}67(3).

\bibitem{Bourouiba2007}
Bourouiba~L, Bartello~P. The intermediate rossby number range and
  two-dimensional–three-dimensional transfers in rotating decaying
  homogeneous turbulence. Journal of Fluid Mechanics.
  2007;\hspace{0pt}587:139--161.

\bibitem{VanBokhoven2008}
Van~Bokhoven~LJA, Cambon~C, Liechtenstein~L, et~al. Refined vorticity
  statistics of decaying rotating three-dimensional turbulence. Journal of
  Turbulence. 2008;\hspace{0pt}9:N6.

\bibitem{Morize2005}
Morize~C, Moisy~F, Rabaud~M. Decaying grid-generated turbulence in a rotating
  tank. Physics of fluids. 2005;\hspace{0pt}17(9):095105.

\bibitem{Morize2006}
Morize~C, Moisy~F. Energy decay of rotating turbulence with confinement
  effects. Physics of Fluids. 2006;\hspace{0pt}18(6):065107.

\bibitem{Praud2006}
Praud~O, Sommeria~J, Fincham~AM. Decaying grid turbulence in a rotating
  stratified fluid. Journal of Fluid Mechanics. 2006;\hspace{0pt}547:389--412.

\bibitem{Godeferd1999}
Godeferd~FS, Lollini~L. Direct numerical simulations of turbulence with
  confinement and rotation. Journal of Fluid Mechanics.
  1999;\hspace{0pt}393:257--308.

\bibitem{Smith2005}
Smith~LM, Lee~Y. On near resonances and symmetry breaking in forced rotating
  flows at moderate rossby number. Journal of Fluid Mechanics.
  2005;\hspace{0pt}535:111--142.

\bibitem{Gallet2014}
Gallet~B, Campagne~A, Cortet~PP, et~al. Scale-dependent cyclone-anticyclone
  asymmetry in a forced rotating turbulence experiment. Physics of Fluids.
  2014;\hspace{0pt}26(3):035108.

\bibitem{Cho2001}
Cho~JYN, Lindborg~E. Horizontal velocity structure functions in the upper
  troposphere and lower stratosphere: 1. observations. Journal of Geophysical
  Research: Atmospheres. 2001;\hspace{0pt}106(D10):10223--10232.

\bibitem{Hakim2005}
Hakim~GJ, Canavan~AK. Observed cyclone–anticyclone tropopause vortex
  asymmetries. Journal of the atmospheric sciences.
  2005;\hspace{0pt}62(1):231--240.

\bibitem{cheng2015laboratory}
Cheng~JS, Stellmach~S, Ribeiro~A, et~al. Laboratory-numerical models of rapidly
  rotating convection in planetary cores. Geophysical Journal International.
  2015;\hspace{0pt}201(1):1--17.

\bibitem{guervilly2014large}
Guervilly~C, Huges~DW, Jones~CA. Large-scale vortices in rapidly rotating
  rayleigh-b{\'e}nard convection. Journal of Fluid Mechanics.
  2014;\hspace{0pt}758:407--435.

\bibitem{vorobieff1998vortex}
Vorobieff~P, Ecke~RE. Vortex structure in rotating rayleigh-benard convection.
  Physica D: Nonlinear Phenomena. 1998;\hspace{0pt}123(1-4):153--160.

\bibitem{Deusebio2014}
Deusebio~E, Boffetta~G, Lindborg~E, et~al. Dimensional transition in rotating
  turbulence. Physical Review E. 2014;\hspace{0pt}90(2):023005.

\bibitem{Naso2015}
Naso~A. Cyclone-anticyclone asymmetry and alignment statistics in homogeneous
  rotating turbulence. Physics of Fluids. 2015;\hspace{0pt}27(3):035108.

\bibitem{Gence2001}
Gence~JN, Frick~C. Naissance des corrélations triples de vorticité dans une
  turbulence statistiquement homogène soumise à une rotation. Comptes Rendus
  de l'Académie des Sciences-Series IIB-Mechanics.
  2001;\hspace{0pt}329(5):351--356.

\bibitem{Sreenivasan2008}
Sreenivasan~B, Davidson~PA. On the formation of cyclones and anticyclones in a
  rotating fluid. Physics of Fluids. 2008;\hspace{0pt}20(8):085104.

\bibitem{Ferrero2009}
Ferrero~E, Mortarini~L, Manfrin~M, et~al. Boundary-layer stress instabilities
  in neutral, rotating turbulent flows. Boundary-layer meteorology.
  2009;\hspace{0pt}130(3):347.

\bibitem{Taylor2010}
Taylor~ZJ, Gurka~R, Kopp~GA, et~al. Long-duration time-resolved piv to study
  unsteady aerodynamics. IEEE Transactions on Instrumentation and Measurement.
  2010;\hspace{0pt}59(12):3262--3269.

\bibitem{jacquin1990homogeneous}
Jacquin~L, Leuchter~O, Cambonxs~C, et~al. Homogeneous turbulence in the
  presence of rotation. Journal of Fluid Mechanics. 1990;\hspace{0pt}220:1--52.

\end{thebibliography}
%%%%%%%%%%%%%%%%%%%%%%%%%%%%%%%%%%%%%%%%%%%%%%%%%%%%%%%%%%%%%%%%%

\end{document}